\documentclass[11pt]{article}
\usepackage{amsmath,amsfonts,amssymb,latexsym}

\def\be{\begin{equation}}
\def\ee{\end{equation}}
\def\bq{\begin{eqnarray}}
\def\eq{\end{eqnarray}}
\def\beq{\begin{eqnarray*}}
\def\eeq{\end{eqnarray*}}

\begin{document}
\title{\Huge{Asymptotic states of generalized universes with higher derivatives}}
\author{\Large{ Spiros Cotsakis\footnote{\texttt{email:\thinspace
skot@aegean.gr}}, Georgios Kolionis\footnote{\texttt{email:\thinspace gkolionis@aegean.gr}}\,
and Antonios Tsokaros\footnote{\texttt{email:\thinspace atsok@aegean.gr}}} \\
{\normalsize Research Group of Geometry, Dynamical Systems and
Cosmology}\\  {\normalsize University
of the Aegean}\\ {\normalsize Karlovassi 83 200, Samos,
Greece}}
\maketitle
\begin{abstract}
\noindent We review ongoing research related to the asymptotic dynamics of isotropic universes in theories with higher derivatives, especially near the initial singularity. We treat two major cases, that is universes in vacuum, and also those filled with radiation using the method of asymptotic splittings of vector fields. Our solutions have the form of formal Fuchsian-type series for the basic unknowns and are valid near collapse or big rip regimes asymptotically.
\end{abstract}

\noindent Isotropic universes in vacuum or those filled with radiation have always been primary examples in efforts to model the effects of higher derivative terms near extreme conditions and cosmological singularities, cf. Refs. \cite{ba-ot}, \cite{RR}, \cite{star}. With the advent of newer and more powerful mathematical methods to study the  asymptotic structure of dynamical systems of Refs.  \cite{CB,me13a}, there is now the prospect of understanding the possible effects near spacetime singularities in a deeper and more complete and transparent way.

Radiation-filled universes in gravity theories with higher derivatives have been analyzed in this way in Refs. \cite{K1,kolionis1}, in an effort to understand their dynamics near the initial cosmological singularity. The
 general higher order action is
\begin{equation}
\mathcal{S}=\frac{1}{2}\int_{\mathcal{M}^4}\mathcal{L}_{\textrm{total}}d\mu_{g},
\end{equation}
where $\mathcal{L}_{\textrm{total}}$ is the lagrangian density of the general
quadratic gravity theory given in the form $\mathcal{L}_{\textrm{total}}=\mathcal{L}(R)+\mathcal{L}_{\textrm{matter}}$, with
\begin{equation}
\mathcal{L}(R)=R + \beta R^2 + \gamma \textrm{Ric}^2 + \delta \textrm{Riem}^2 ,
\label{eq:lagra}
\end{equation}
where $\beta,\gamma,\delta$ are constants.
The field equations  derived are as follows:
\begin{equation}
R^{\mu\nu}-\frac{1}{2}g^{\mu\nu}R+
      \frac{\xi}{6} \left[2RR^{\mu\nu}-\frac{1}{2}R^2g^{\mu\nu}-2(g^{\mu\rho}g^{\nu\sigma}-g^{\mu\nu}g^{\rho\sigma})\nabla_{\rho}\nabla_{\sigma}R \right]=T^{\mu\nu},
\label{eq:fe}
\end{equation}
where we have set  $\xi=2(3\beta+\gamma+\delta)$. This naturally splits into
$00$- and $ii$-components ($i=1,2,3$), but only the $00$-component of
(\ref{eq:fe}) will be used below.
Using  the standard FRW metric, the field equation (\ref{eq:fe}) leads to our basic cosmological equation  in the form
\begin{equation}
\frac{k+\dot{a}^2}{a^2}+\xi\left[2\: \frac{\dddot{a}\:\dot{a}}{a^2} + 2\:\frac{\ddot{a}\dot{a}^2}{a^3}-\frac{\ddot{a}^2}{a^2} - 3\:
\frac{\dot{a}^4}{a^4} -2k\frac{\dot{a}^2}{a^4} + \frac{k^2}{a^4}\right] = \frac{\zeta^2}{a^4} ,
\label{eq:beq}
\end{equation}
where $\zeta$ is a constant defined by the constraint
\begin{equation}
\frac{\rho}{3}=\frac{\zeta^2}{a^4},\quad (\textrm{from}\,\,\nabla_{\mu}T^{\mu 0}=0).
\end{equation}
and $k$ is the (constant) curvature normalized to take the three values $0, +1$ or $-1$ for the complete, simply connected,  flat, closed or open space sections.

Setting $x=a$, $y=\dot{a}$ and $z=\ddot{a}$, Eq. (\ref{eq:beq}) can be written as an autonomous dynamical
system of the form
\begin{equation} \label{basic dynamical system}
\mathbf{\dot{x}}=\mathbf{f}_{\,k,\textsc{RAD}}(\mathbf{x}),\quad \mathbf{x}=(x,y,z),
\end{equation}
where  the \emph{curvature-radiation} vector field $\mathbf{f}_{\,k,\textsc{RAD}}:\mathbb{R}^3\rightarrow\mathbb{R}^3:(x,y,z)\mapsto\mathbf{f}_{\,k,\textsc{RAD}}(x,y,z)$ that defines the system is given by the form
\begin{equation} \label{vf}
\mathbf{f}_{\,k,\textsc{RAD}}(x,y,z)=\left( x,y,\frac{\zeta^2-k^2\xi}{2\xi x^2y} + \frac{3y^3}{2x^2} + \frac{z^2}{2y} -\frac{yz}{x} - \frac{y}{2\xi}
-\frac{k}{2\xi y} + \frac{ky}{x^2}\right).
\end{equation}
The curvature-radiation field $\mathbf{f}_{\,k,\textsc{RAD}}$ combines the effects of curvature and radiation and describes completely the dynamical evolution of any radiation-filled FRW universe in higher order gravity. Following the asymptotic methods of Ref. \cite{CB}, we conclude that the only acceptable asymptotic splitting of the vector field  $\mathbf{f}^{(0)}_{\,k,\textsc{RAD}}$ is of the form
\begin{equation}
\mathbf{f}_{\,k,\textsc{RAD}}=\mathbf{f}^{(0)}_{\,k,\textsc{RAD}} + \mathbf{f}^{\,(\textrm{sub})}_{\,k,\textsc{RAD}},
\end{equation}
with with dominant part
\begin{equation}
\mathbf{f}^{(0)}_{\,k,\textsc{RAD}}(\mathbf{x})=\left(y,z,\frac{3y^3}{2x^2} + \frac{z^2}{2y} -\frac{yz}{x}
\right), \label{eq:f0}
\end{equation}
and subdominant part
\begin{equation} \label{eq:f01}
\mathbf{f}^{\,(\textrm{sub})}_{\,k,\textsc{RAD}}(\mathbf{x})=
\left(0,0, \frac{\zeta^2-k^2\xi}{2\xi x^2y} - \frac{y}{2\xi} -\frac{k}{2\xi y} + \frac{ky}{x^2}\right).
\end{equation}
We then look for the possible asymptotic solutions, asymptotic forms of integral curves  of the curvature-radiation field $\mathbf{f}_{\,k,\textsc{RAD}}$, that is we search for the dominant balances determined by the dominant part $\mathbf{f}^{(0)}_{\,k,\textsc{RAD}}$ given by Eq. (\ref{eq:f0}). For this purpose, we substitute in the dominant system $(\dot x,\dot y,\dot z)(t)=\mathbf{f}^{(0)}_{\,k,\textsc{RAD}}$ the forms $\mathbf{x}(t)=\mathbf{a}t^{\mathbf{p}}=(\theta t^{p}, \eta
t^{q}, \rho t^{r})$ and solve the resulting nonlinear algebraic system to determine  the dominant balance $(\mathbf{a},\mathbf{p})$ as an exact, scale invariant solution. This leads to the unique \emph{curvature-radiation} balance $\mathcal{B}_{\,k,\textsc{RAD}}\in\mathbb{C}^3\times\mathbb{Q}^3$,  with
\begin{equation}   \label{basic balance}
\mathcal{B}_{\,k,\textsc{RAD}}=(\mathbf{a},\mathbf{p})= \left(\left(
\theta,\frac{\theta}{2},-\frac{\theta}{4}\right),\:
\left(\frac{1}{2},-\frac{1}{2},-\frac{3}{2}\right)\right),
\end{equation}
where $\theta$ is a real, arbitrary constant. In particular, this means that the vector field $\mathbf{f}^{(0)}_{\,k,\textsc{RAD}}$ is \emph{a scale invariant system}.

After calculating the spectrum of the Kovalevskaya matrix which is of the form
\begin{equation}
\textrm{spec}(\mathcal{K}_{\,k,\textsc{RAD}})=\{-1,0,3/2\},
\end{equation}
in order to obtain the the $\mathcal{K}$-exponents of the dominant system, we proceed with the construction of the final solution by substituting the forms
\begin{equation} \label{eq:series}
x(t) = \sum_{i=0}^{\infty} c_{1i} t^{\frac{i}{2}+\frac{1}{2}}, \:\:\:\:\:
y(t) = \sum_{i=0}^{\infty} c_{2i}  t^{\frac{i}{2}-\frac{1}{2}},\:\:\:\:\:
z(t) = \sum_{i=0}^{\infty} c_{3i} t^{\frac{i}{2}-\frac{3}{2}} ,\:\:\:\:\:
\end{equation}
where  $c_{10}=\theta ,c_{20}=\theta /2,c_{30}=-\theta
/4$, and we are led to various recursion relations that determine the unknowns $c_{1i}, c_{2i}, c_{3i}$ term by term. Further algebraic manipulations lead to the final series representation of the solution in the form:
\begin{equation}
x(t) = \theta \:\:t^{1/2} - \frac{k}{2\theta}\:\:t^{3/2} +
c_{13} \:\: t^{2} + \displaystyle
\left(\frac{4\zeta^2-\theta^4}{12\xi\theta^3}-\frac{k^2}{8\theta^3}\right)\:\:t^{5/2} + \cdots .
\label{eq:gensol}
\end{equation}
Our series (\ref{eq:gensol}) has three arbitrary constants, $\theta,
c_{13}$ and a third one corresponding to the arbitrary position of
the singularity (taken here to be zero without loss of generality), and is therefore a local expansion of the
\emph{general} solution around the initial singularity. Since
the leading order coefficients are real,  we conclude that there is an open set of
 initial conditions  for which the general solution blows up at the
finite time (initial) singularity at $t=0$.  This proves the
stability of our solutions in the neighborhood of the singularity.

The case of vacuum universes with higher derivatives is currently under investigation, cf. Ref. \cite{kolionis2}, both for the flat and the curved subcases. The dynamics of the various families of models depends on the asymptotic modes of the \emph{curvature-vacuum} vector field
\begin{equation}
\mathbf{f}_{\,k,\textsc{VAC}}(x,y,z)=\left( y,\frac{y^{2}}{2x}-3xy+kxz-\frac{k^2 z^2}{2x}- \frac{x}{12\epsilon}-\frac{kz}{12\epsilon x},-2xz \right),
\end{equation}
and this leads to very interesting, possible dynamical regimes asymptotically. In the flat-vacuum case, we can show that there is a general solution of the field equations with the scale factor having the leading order behaviour $t^{1/2}$. In the curved cases, the situation becomes more involved and there can be various general and particular solution of great interest. However, it appears that the `radiation-vacuum' solution mentioned above is a global attractor of the possible curved asymptotics. These results will be reported in detail elsewhere.


\begin{thebibliography}{99}

\bibitem{ba-ot}
J. D. Barrow and A. C. Ottewill, \emph{J. Phys.} \textbf{A16} (1983) 2757.
\bibitem{RR} T. V. Ruzmaikina and A. A. Ruzmaikin, {\it Zh. Eksp. Teor. Fiz.} {\bf 57} (1970) 680.
\bibitem{star} A. A. Starobinski, \emph{Phys. Lett.} \textbf{B91} (1980) 99.
\bibitem{CB}
S. Cotsakis and J. D. Barrow, \emph{J. Phys. Conf. Ser.} \textbf{68} (2007) 012004.
\bibitem{me13a}
S. Cotsakis, \emph{Asymptotic Poincar\'e compactification and finite-time singularities}, arXiv: 1301.4778.
\bibitem{K1}
S. Cotsakis and A. Tsokaros, Phys. Lett. \textbf{B651} (2007) 341-344.
\bibitem{kolionis1}
S. Cotsakis, G. Kolionis and A. Tsokaros, \emph{The initial state of generalized radiation universes}, arXiv:1211.5255.
\bibitem{kolionis2}
S. Cotsakis, G. Kolionis and A. Tsokaros, (in preparation).
\end{thebibliography}
\end{document}